\documentclass[reprint,aps,prb,graphicx,superscriptaddress]{revtex4-2}
\usepackage{graphicx}
\usepackage{xcolor}
\usepackage[hidelinks]{hyperref}
\hypersetup{
    colorlinks,
    linkcolor={blue},
    citecolor={blue},
    urlcolor={blue}
}

\usepackage{afterpage}

\graphicspath{{figures/}}

\draft 

\newcommand{\CsVSb}{CsV$_3$Sb$_5$}

\begin{document}
\renewcommand{\figurename}{Fig.}


\title{Evidence for reduced periodic lattice distortion within \\ the Sb-terminated surface layer of the kagome metal \CsVSb{}}

\author{Felix Kurtz}
\affiliation{Max Planck Institute for Multidisciplinary Sciences, 37077 Göttingen, Germany}
\affiliation{4th Physical Institute -- Solids and Nanostructures, University of Göttingen, 37077 Göttingen, Germany}

\author{Gevin von Witte}
\affiliation{Max Planck Institute for Multidisciplinary Sciences, 37077 Göttingen, Germany}
\affiliation{Institute for Biomedical Engineering, University and ETH Zurich, 8092 Zurich, Switzerland}
\affiliation{Institute of Molecular Physical Science, ETH Zurich, 8093 Zurich, Switzerland}

\author{Lukas Jehn}
\affiliation{Max Planck Institute for Multidisciplinary Sciences, 37077 Göttingen, Germany}
\affiliation{4th Physical Institute -- Solids and Nanostructures, University of Göttingen, 37077 Göttingen, Germany}

\author{Alp Akbiyik}
\affiliation{Max Planck Institute for Multidisciplinary Sciences, 37077 Göttingen, Germany}
\affiliation{4th Physical Institute -- Solids and Nanostructures, University of Göttingen, 37077 Göttingen, Germany}

\author{Igor Vinograd}
\affiliation{Max Planck Institute for Multidisciplinary Sciences, 37077 Göttingen, Germany}
\affiliation{4th Physical Institute -- Solids and Nanostructures, University of Göttingen, 37077 Göttingen, Germany}

\author{Matthieu Le Tacon}
\author{Amir A. Haghighirad}
\affiliation{Institute for Quantum Materials and Technologies, Karlsruhe Institute of Technology, 76021 Karlsruhe, Germany}

\author{Dong Chen}
\author{Chandra Shekhar}
\author{Claudia Felser}
\affiliation{Max Planck Institute for Chemical Physics of Solids, 01187 Dresden, Germany}

\author{Claus Ropers}
\email[Author to whom correspondence should be addressed: ]{claus.ropers@mpinat.mpg.de}
\affiliation{Max Planck Institute for Multidisciplinary Sciences, 37077 Göttingen, Germany}
\affiliation{4th Physical Institute -- Solids and Nanostructures, University of Göttingen, 37077 Göttingen, Germany}

\date{\today}

\begin{abstract}
The discovery of the kagome metal \CsVSb{} sparked broad interest, due to the coexistence of a charge density wave (CDW) phase and possible unconventional superconductivity in the material. In this study, we use low-energy electron diffraction (LEED) with a µm-sized electron beam to explore the periodic lattice distortion at the antimony-terminated surface in the CDW phase. We recorded high-quality backscattering diffraction patterns in ultrahigh vacuum from multiple cleaved samples. Unexpectedly, we did not find superstructure reflexes at intensity levels predicted from dynamical LEED calculations for the reported $2 \times 2 \times 2$ bulk structure. Our results suggest that in \CsVSb{} the periodic lattice distortion accompanying the CDW is less pronounced at Sb-terminated surfaces than in the bulk.
\end{abstract}

\maketitle

\begin{figure*}[t]
    \centering
    \includegraphics{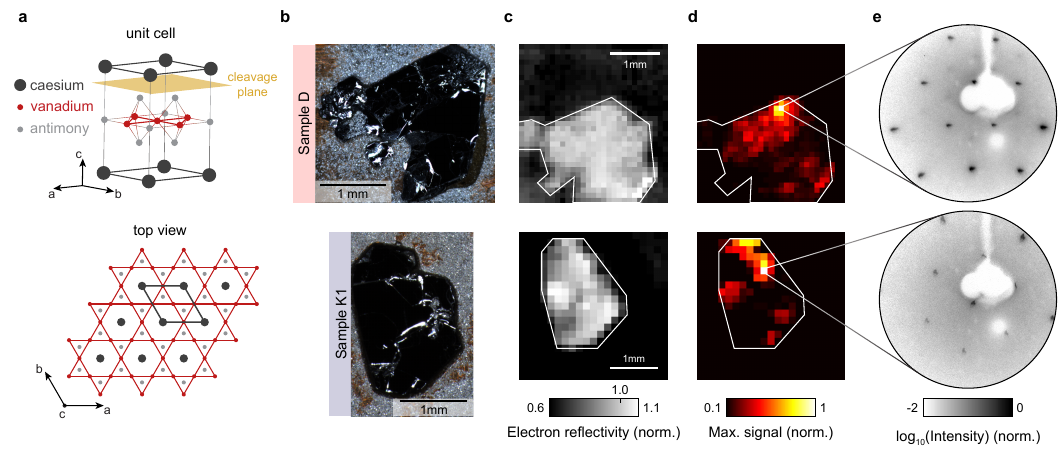}
    \caption{
        \textbf{a,} Crystal structure of \CsVSb{} showcasing a vanadium kagome net. The cleavage plane between caesium and antimony planes is indicated.
        \textbf{b,} Optical microscope images of two \CsVSb{} samples.
        \textbf{c,} Corresponding spatial scans where each pixel represents the total intensity in the recorded diffraction image (integration time 3\,s), resembling the shape of the sample.
        \textbf{d,} Same spatial scans where each pixel displays the maximum intensity in the recorded diffraction image, indicating positions of high surface quality and low local curvature resulting in bright and sharp diffraction spots.
        \textbf{e,} Diffraction images taken at the areas with high total and maximum intensity (90\,eV electron energy, integration times of 20\,s in the top image and 30\,s in the bottom image, 30\,K sample temperature).
    }
    \label{fig:Overview}
\end{figure*}


A new family of superconducting kagome materials has recently attracted much attention due to the rich phenomenology they offer, including a band structure containing a flat band, a Dirac cone and multiple van Hove singularities \cite{ortizNewKagomePrototype2019, wilsonAV3Sb5KagomeSuperconductors2024}. The vanadium-based kagome materials \textit{A}V$_3$Sb$_5$ with \textit{A} = (Rb, K, Cs) exhibit superconductivity at ambient pressure below temperatures $T_{\textnormal{c}} \approx$ 0.75\,K, 0.93\,K and 2.5\,K, respectively \cite{yinSuperconductivityNormalStateProperties2021, ortizSuperconductivityZ2Kagome2021, ortizCsV3Sb5Z2Topological2020}. Remarkably, the superconducting phase emerges within a charge-ordered state with a 2 $\times$ 2 motif of the charge density wave (CDW). The CDW in the kagome layers onsets at much higher temperatures of $T_{\mathrm{CDW}} = 78-103$\,K. Significant effort has been devoted to uncovering the mechanism behind CDW formation. Band structure calculations indicate that strong electron-phonon coupling renders the lattice unstable \cite{christensenTheoryChargeDensity2021}. However, a phonon softening at the CDW wave vector, as expected for a Peierls-like mechanism, was not observed experimentally \cite{liObservationUnconventionalCharge2021}. An electron-driven nesting of the high density of states at saddle points located at the $M$ points on the Brillouin zone boundary would lead to the observed 2 $\times$ 2 motif. However, 2D Fermi surface nesting as a driving mechanism of the CDW has been questioned \cite{liDiscoveryConjoinedCharge2022,kaboudvandFermiSurfaceNesting2022}, rather accentuating the inter-layer coupling which leads to out-of-plane stacking with $2 \times 2 \times 2$, or $2 \times 2 \times 4$ supercells \cite{ortizFermiSurfaceMapping2021,stahlTemperaturedrivenReorganizationElectronic2022,kautzschStructuralEvolutionKagome2023,frassinetiMicroscopicNatureChargedensity2023,wangStructureKagomeSuperconductor2023,xiaoCoexistenceMultipleStacking2023,plumbPhaseSeparatedChargeOrder2024}. 

As with superconductivity \cite{mielkeTimereversalSymmetrybreakingCharge2022}, there are reports of an unconventional nature of the CDW, suggesting a state with broken time-reversal symmetry (TRS) \cite{mielkeTimereversalSymmetrybreakingCharge2022} despite the absence of static magnetism \cite{ortizNewKagomePrototype2019}. This may arise from a complex CDW order parameter that induces orbital currents in the kagome layers \cite{christensenLoopCurrentsAV3Sb52022}. Early scanning tunneling microscopy (STM) and resistivity measurements also observed chirality as well as nematicity in the CDW state \cite{jiangUnconventionalChiralCharge2021, shumiyaIntrinsicNatureChiral2021,xiangTwofoldSymmetryCaxis2021} -- effects rarely seen in other CDW materials and suggesting an alternative electronic mechanism that is derived from the unusual kagome band structure \cite{kieselUnconventionalFermiSurface2013}. However, it remains unresolved whether TRS breaking \cite{saykinHighResolutionPolar2023}, orbital currents \cite{liegeSearchOrbitalMagnetism2024}, chirality \cite{liNoObservationChiral2022, elmersChiralityKagomeMetal2024} or nematicity \cite{frachetColossalCAxisResponse2024} indeed occur in these kagome materials. Some of the observed discrepancies may result from intrinsic differences in the surface and bulk properties. Whereas optical conductivity and Raman scattering found CDW gap energies between 80 and 100\,meV \cite{uykurLowenergyOpticalProperties2021, zhouOriginChargeDensity2021, heAnharmonicStrongcouplingEffects2024}, gaps derived from STM and ARPES vary from 20\,meV \cite{kangTwofoldVanHove2022,liangThreeDimensionalChargeDensity2021} to 50-70\,meV \cite{nakayamaMultipleEnergyScales2021}. Important insights on the surface properties were gained by STM studies that concentrated on the Sb-terminated surface due to its robustness, and which uncovered an additional unidirectional $1 \times 4$ stripe order \cite{chenRotonPairDensity2021,zhaoCascadeCorrelatedElectron2021,wangElectronicNatureChiral2021,shumiyaIntrinsicNatureChiral2021}. To date, the associated structural properties are not known, which calls for crystallographic investigations using surface-sensitive techniques.

In this work, we use low-energy electron diffraction (LEED) with a µm-sized electron beam on \CsVSb{} in its CDW phase. We find no diffraction signature of a CDW-coupled periodic lattice distortion (PLD) on the Sb-terminated surface, while dynamical LEED simulations for the x-ray-determined $2 \times 2 \times 2$ bulk structure predict superstructure peaks well above the noise level in our experiment. The robustness of this observation is tested on several samples from batches of crystals grown at two different institutions. Our observations suggest a reduced lattice distortion at the Sb-terminated surface and highlight the importance of surface-sensitive structural probes to complement the interpretation of electronic properties measured by ARPES or STM.  

The material's unit cell consists of four atomic layers, namely a strongly-bound V$_3$Sb$_5$ stack sandwiched between Cs layers (Fig.\,\ref{fig:Overview}a), with the vanadium atoms arranged in a kagome net. Antimony atoms fill the inner sites of the hexagons and additionally arrange in a honeycomb lattice above and below the kagome layer. As a consequence of the weaker bonding, the cleavage plane is between Cs and Sb planes, creating two possible terminations with either caesium or antimony at the surface.

In the experiments, cleaving multiple samples usually resulted in only small areas of high crystalline quality, and we found it necessary to study the surface with a µm-scale electron beam. In particular, we employed our home-built electron gun with a beam diameter of 80-100\,µm \cite{vogelgesangPhaseOrderingCharge2018}. We screened 6 cleaves of 4 different crystals for superstructure reflexes. On two cleaved samples, shown in Fig.\,\ref{fig:Overview}b, we performed spatial LEED scans. In these measurements, LEED images with an integration time of 3\,s were taken across a square grid on the entire sample (pixel spacing 100\,µm and 150\,µm, respectively). For every scan position, we display the total reflected electron signal (Fig.\,\ref{fig:Overview}c) and the maximum intensity within the diffraction image (Fig.\,\ref{fig:Overview}d), which indicates the quality and the sharpness of the position-dependent diffraction pattern. Overall, the macroscopic shapes of the crystals are represented by an elevated, comparably homogeneous total electron reflectivity. Yet, only small areas, roughly 200\,µm in diameter, yield sharp diffraction images with high peak intensities that imply high surface quality and flatness over the electron beam diameter. Example diffraction images of these high-quality locations are shown in Fig.\,\ref{fig:Overview}e. Both images are taken with electrons at 90\,eV energy, and at a sample temperature of 30\,K, significantly below the phase transition temperature, $T_{\mathrm{CDW}} = 94$\,K. Nonetheless, under these conditions, no evidence for superstructure diffraction peaks is found. 

\begin{figure}[!htb]
    \centering
    \includegraphics{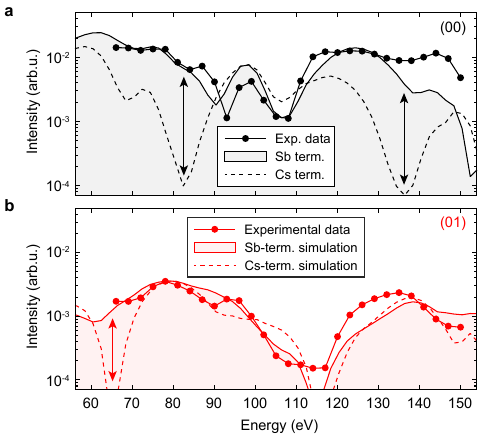}
    \caption{Experimental spot intensities compared with dynamical LEED calculations, considering the (00) spot in (a) and the (01) spots in (b). Simulations assume either a complete antimony or caesium termination. We find good agreement with an assumed antimony termination, and several spot extinctions expected for caesium termination are not observed (arrows).}
    \label{fig:theo_vs_exp}
\end{figure}

We cleave the crystals at room temperature before cooling, which can lead to two distinct terminations \cite{chenRotonPairDensity2021}, namely a $1 \times 1$ antimony termination or a partial $\sqrt{3} \times \sqrt{3}$ caesium termination. Similar observations were made for RbV$_3$Sb$_5$ \cite{yuEvolutionElectronicStructure2022}. Some signatures of partial caesium termination are visible by additional spots, faintly seen in the top panel of Fig.\,\ref{fig:Overview}e. Yet we note that these $(\sqrt{3} \times \sqrt{3})R30^\circ$ peaks are either very weak or not visible at all, depending on the probed position, suggesting that this termination covers a rather small fraction of the total surface area. Furthermore, these spots disappear when the sample is held at room temperature for several hours, likely due to increased Cs desorption (see Fig.\,\ref{fig:disappearance_sqrt(3)_sqrt(3)} in the Appendix). Overall, these observations indicate that we are dealing primarily with Sb terminations in our measurements. This is further corroborated by electron energy dependent measurements in conjecture with dynamical LEED calculations below (Fig.\,\ref{fig:theo_vs_exp}).

\begin{figure}[!b]
    \centering
    \includegraphics{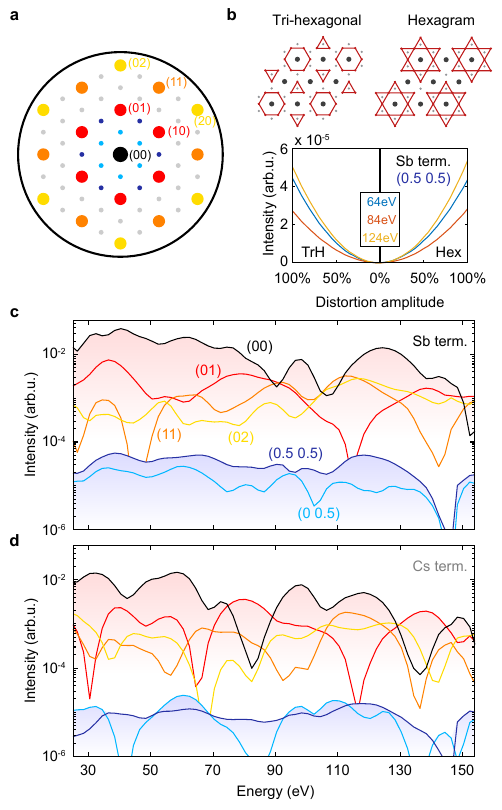}
    \caption{Dynamical LEED calculations:
        \textbf{a,} Schematic LEED pattern with strongest expected superstructure peaks in blue.
        \textbf{b,} Quadratic scaling of the diffraction signal with the distortion amplitude, shown here for the Sb termination, the (0.5 0.5) spot and three different energies. The left side shows the distortion towards the tri-hexagonal structure, whereas the right side treats the hexagram distortion.
        \textbf{c, d,} Energy-dependent signal strength of the four strongest main lattice peaks as well as the two strongest CDW peaks for both Cs and Sb terminations, retrieved from dynamical LEED calculations. The distortions are taken from Ref. \cite{stahlTemperaturedrivenReorganizationElectronic2022} (Tri-hexagonal).}
    \label{fig:Dynamical_LEED}
\end{figure}

In order to obtain quantitative predictions of expected superstructure peak intensities associated with the CDW, we perform dynamical LEED calculations in the electron energy range from 25 to 155\,eV. In these computations, we assume the $2 \times 2 \times 2$ lattice distortions as obtained from x-ray diffraction in the bulk \cite{stahlTemperaturedrivenReorganizationElectronic2022}. Two distinct structural distortions are discussed in the literature: the hexagram distortion (also known as Star-of-David) and its inverse, the tri-hexagonal distortion (Fig.\,\ref{fig:Dynamical_LEED}b, top). Both distortions produce the $2 \times 2$ motif in the kagome plane, while staggering these motifs results in a doubling of the unit cell along the vertical direction. The distortion amplitudes for both cases are estimated from the same bulk diffraction data, and thus also show large similarities in our calculations. For simplicity, in the following, we only display the curves for the tri-hexagonal distortion, which represents the suggested ground state by density functional theory calculations \cite{miaoGeometryChargeDensity2021,tanChargeDensityWaves2021}. The results for the hexagram distortion are shown in Fig.\,\ref{fig:dynLEED_SoD_all} in the Appendix. In Fig.\,\ref{fig:Dynamical_LEED}a, the PLD spots are indicated, and the sixfold symmetry becomes apparent. The energy-dependent spot intensities are displayed in Figs.\,\ref{fig:Dynamical_LEED}c and \ref{fig:Dynamical_LEED}d for complete Sb or Cs surface terminations, respectively. We show the specular reflex and the main-lattice peaks of first and second order. Furthermore, the two strongest superstructure spots are shown, which are generally between two and three orders of magnitude less intense than the main spots. In particular, the (0.5 0.5) spot appears nearly energy-independent in the range of 40 to 130\,eV, regardless of termination.

Before applying the presented dynamical LEED calculations to the search for superstructure reflexes, we first use them to determine the surface termination in the experiment. Namely, we perform LEED-\textit{I}(V) measurements, i.e. we record spot intensities as a function of electron energy, on the high-quality surface area of sample D and compare them with the predictions made for either a complete antimony or caesium termination (Fig.\,\ref{fig:theo_vs_exp}). Experimentally, we only analyze the spots that are visible throughout the entire energy scan from 66\,eV to 150\,eV, namely three (01) spots, over which we average, and the (00) spot. Indeed, we find a fairly good agreement with the Sb-terminated simulation as proposed above, while an assumed Cs termination predicts several electron energies (indicated by arrows) at which the spots should disappear -- in stark contrast to the experimental data. Combined with the general weakness of the $(\sqrt{3} \times \sqrt{3})R30^\circ$ spots discussed above, we find predominantly antimony-terminated surfaces in our experiments.

\begin{figure}[!b]
    \centering
    \includegraphics{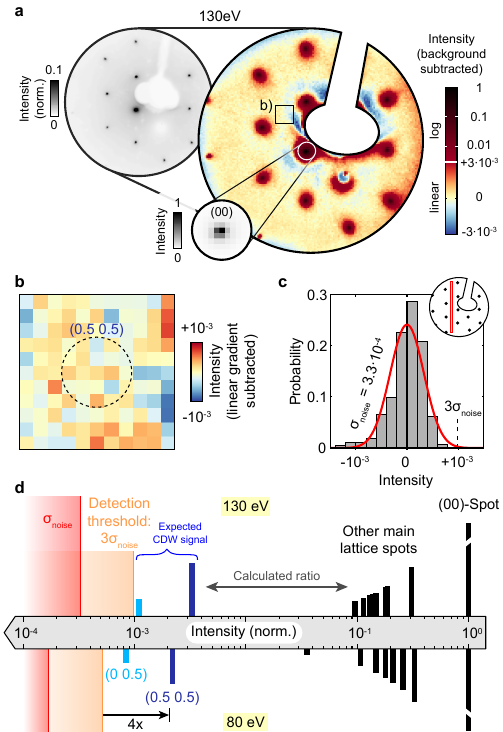}
    \caption{Search for superstructure peaks and noise analysis:
        \textbf{a,} LEED image at 130\,eV and with an integration time of 11 minutes, normalized to the (00) spot (sample K2). Left image has a linear color scale capped at 0.1, highlighting the momentum resolution and the sharp diffraction spots. Right image illustrates the low noise floor after binning and the subtraction of a slowly-varying background. Zoom-in displays the (00) spot with a linear color scale.
        \textbf{b,} Region around (0.5 0.5), where a superstructure spot is expected. A linear intensity gradient was subtracted.
        \textbf{c,} Histogram of the intensity within the red rectangle shown in the schematic diffraction pattern. The standard deviation $\sigma_{\text{noise}}$ is $3.3\cdot 10^{-4}$, and the detection threshold for a peak is defined by $3\sigma_{\text{noise}}$.
        \textbf{d,} Experimental intensities of the main lattice spots as well as expected strength of the CDW peaks, which is well above the detection threshold. The same analysis is done for a corresponding diffraction image at 80\,eV and displayed in the lower half of the panel.}
    \label{fig:Absence_CDW_130eV}
\end{figure}

After screening several surfaces at various electron energies and finding no indication of superstructure spots, we enhance the signal-to-noise ratio by prolonging the integration time. Based on the predictions of the dynamical LEED calculations, we record diffraction images at a high-quality surface position of sample K2. Using an electron energy of 130\,eV, we increase the dynamic range of the measurement by integrating multiple camera frames over an extended duration of 11 minutes (Fig.\,\ref{fig:Absence_CDW_130eV}a, left). At this electron energy, first- and second-order main lattice spots appear at a similar intensity, and multiple locations of potential superstructure peaks are accessible in the diffraction pattern. We now analyze these high-quality backscattering diffraction images, focusing on the search for superstructure peaks. To this end, it is necessary to determine the noise floor. After normalizing with a flat-field image and subtracting a slowly varying background (see Appendix \ref{sec:image_analysis} for further details), we obtain the image shown on the right in Fig.\,\ref{fig:Absence_CDW_130eV}a. To further increase the signal-to-noise ratio, we binned the image considerably such that the observed main-lattice spots cover only a few pixels, as shown in the zoom-in of the (00) spot. Although this substantially increases the sensitivity to faint diffraction features, the measurements do not exhibit signs of a superstructure peak exceeding the remaining noise level. As an example, we depict the region around a (0.5 0.5) location in Fig.\,\ref{fig:Absence_CDW_130eV}b, where a linear intensity gradient within the shown square is subtracted. In order to estimate an upper boundary for the intensity of possibly undetected superstructure peaks, we determine the background noise level. To this end, we create an intensity histogram within a stripe between the main lattice spots (Fig.\,\ref{fig:Absence_CDW_130eV}c). Its standard deviation is $\sigma_{\text{noise}} = 3.3 \cdot 10^{-4}$. We define the detection threshold as $3\sigma_{\text{noise}} = 9.9 \cdot 10^{-4}$, meaning that, under Gaussian noise, only 1.5 out of every 1000 pixels are statistically expected to exceed this threshold. With the spot position known to a precision of up to $5 \times 5$ pixels at most, the probability of mistakenly attributing a noise-induced count as a diffraction peak is exceedingly low.

We now compare the obtained detection threshold with the expected superstructure intensities from Fig.\,\ref{fig:Dynamical_LEED} by summarizing the different spot intensities in Fig.\,\ref{fig:Absence_CDW_130eV}d. This panel shows the intensities of the (00) and the first- and second-order main lattice spots, which are determined from the experimental data in the LEED image. Furthermore, the expected (0 0.5) and (0.5 0.5) CDW spot intensities from our dynamical LEED calculations are depicted for an Sb-terminated surface. We hereby consider the sum of all observed main lattice peaks rather than a single spot as the reference between experiment and simulation. Furthermore, we present the results of the same analysis performed for a diffraction image at 80\,eV. As can be seen, all predicted intensities of the displayed CDW spots are above the detection threshold of $3\sigma_{\text{noise}}$, regardless of the two displayed electron energies. These spots should therefore be visible in both diffraction images. In particular, at an electron energy of 80\,eV the expected peak strength of the (0.5 0.5) spots is roughly four times larger than our detection threshold, which suggests a reduced lattice distortion. Since the intensity of the superstructure peaks scales quadratically with the lattice distortion (see Fig.\,\ref{fig:Dynamical_LEED}b), we estimate a PLD amplitude within the surface layer that is less than half its bulk value.

We discuss possible factors for a reduced strength of superstructure reflections in the context of the literature and our LEED experiments. Firstly, due to the relaxation of surface bonds, the amplitude of lattice distortions can be reduced with respect to the bulk, and moderate deviations between bulk and surface CDW distortions have been previously observed in TaS$_2$ \cite{vonwitteSurfaceStructureStacking2019}. In comparison, the deduced reduction of the distortions by at least a factor of two in \CsVSb{} appears to be more significant, which could point towards a different formation mechanism in these CDW materials. Secondly, the polarity of the cleaved surface must be considered, which changes depending on whether one is probing a predominantly Sb- or Cs-terminated surface, as the ionic bonds between antimony and alkali metal atoms are broken. Thus, the Sb-terminated surface is effectively hole doped. Recent termination-resolved ARPES studies consistently observed a CDW-induced splitting only on the alkali-terminated surfaces of all three \textit{A}V$_3$Sb$_5$ compounds \cite{katoPolaritydependentChargeDensity2022,katoSurfaceterminationdependentElectronicStates2023}. The findings are explained by the polar nature of the surface leading to energy shifts of key electronic bands. Moreover, this behavior is reminiscent of the asymmetry found between moderate hole or electron doping by Sn- or Te-substitution of antimony in bulk samples \cite{capasalinasElectronholeAsymmetryPhase2023}: while in the latter case, the CDW state is largely preserved, hole doping rapidly suppresses the long-range charge order. 

Our µm-beam LEED measurements imply a suppression of the CDW-coupled lattice distortion on Sb-terminated surfaces. Importantly, however, various STM studies were indeed able to observe a $2 \times 2$ motif on this termination, often superimposed with a uniaxial $1 \times 4$ modulation \cite{chenRotonPairDensity2021, zhaoCascadeCorrelatedElectron2021, wangElectronicNatureChiral2021, shumiyaIntrinsicNatureChiral2021, liangThreeDimensionalChargeDensity2021}. As we did not observe a unidirectional PLD either, the possible structural distortions of both surface modulations remain elusive. One might speculate whether the interplay of these two charge modulations may weaken their influence on atomic displacements. Our findings contribute to the ongoing debate, highlighting at least partially conflicting results that likely arise from the intricate band structure of \CsVSb{}. Taken together, these results emphasize the need for complementary experimental approaches, since electronic and structural manifestations of the CDW, as well as its bulk and surface properties, can differ significantly. Further LEED studies of \CsVSb{} following the adsorption of caesium or different dopants could provide valuable insights to help resolve this puzzle. More generally, micron-scale surface-structural probing may help to elucidate origins of distinct charge-density wave or pair-density wave properties between the surface and the bulk of further correlated materials, such as UTe$_2$\cite{aishwaryaMagneticfieldsensitiveChargeDensity2023, theussAbsenceBulkThermodynamic2024, kengleAbsenceBulkSignature2024} or hole-doped cuprates \cite{edkinsMagneticFieldInduced2019, maggio-aprileVortexcoreSpectroscopyDwave2023, dasilvanetoUbiquitousInterplayCharge2014, vinogradLocallyCommensurateChargedensity2021, blackburnSearchingSignaturePair2023}.

\begin{acknowledgments}
This work was funded by the European Research Council (ERC Advanced Grant “ULEEM,” ID: 101055435) and the Gottfried Wilhelm Leibniz program. F.K. gratefully acknowledges support from the  Max Planck School of Photonics and I.V. acknowledges the Horizon Europe MSCA fellowship 101065694. C.F. acknowledges support from the Deutsche Forschungsgemeinschaft (DFG) under SFB1143 (Project No. 247310070), the Würzburg-Dresden Cluster of Excellence on Complexity and Topology in Quantum Matter—ct.qmat (EXC 2147, Project No. 390858490) and the QUAST-FOR5249-449872909.
\end{acknowledgments}




\appendix
\section{Crystal synthesis}
\label{sec:crystal_synthesis}
The crystals were grown using the self-flux method \cite{ortizCsV3Sb5Z2Topological2020}. Sample D in Fig.\,\ref{fig:Overview} was grown at the Max Planck Institute for Chemical Physics of Solids (batch D), while samples K1 and K2 were grown at the Karlsruhe Institute of Technology (batch K).

\textit{Batch D:}
Cs, V, and Sb with atomic ratio of 7:3:14 were loaded in an alumina crucible, and subsequently sealed in a tantalum crucible and quartz tube in turn. The tube was heated to 1000\,℃, hold for 20\,h, and cooled to 400\,℃ with a rate of 3\,℃/h before the furnace was turned off. The crystals were obtained by dissolving the flux by demineralized water.

\textit{Batch K:} High purity elemental reagents, Cs (Alfa Aesar, 99.98\,\%), V (Cerac/Pure, 99.9\,\% , purified to remove oxygen) and Sb (Gmaterials, 99.999\,\%) were mixed in molar ratio 2:1:6 in an alumina crucible, and this was placed in an iron container, which was welded in argon atmosphere. The loaded container was put in a tube furnace (the tube was sealed in argon atmosphere) and heated to 1050\,°C and soaked there for 15\,h. The material was cooled to 650\,°C at 2\,°C/h and the furnace was canted to flow off the excess flux. The furnace was cooled to room temperature and the crystals were removed from the flux mechanically and the remaining flux on the surface of the crystals was washed by demineralized water. All single crystals have been well characterized prior to the XRD-experiments and energy dispersive x-ray spectroscopy described below. Chemical composition of \CsVSb{} crystals was performed using energy dispersive x-ray spectroscopy (EDS) in a COXEM EM-30plus electron microscope equipped with an Oxford Silicon-Drift-Detector (SDD) and AZtecLiveLite-software package. The EDS revealed a quantitative elemental composition of Cs:V:Sb 1:3:5.17, respectively.

\section{LEED measurements}
\label{sec:LEED}
The experiments are conducted using a recently developed ultrafast low-energy electron diffraction (ULEED) apparatus \cite{vogelgesangPhaseOrderingCharge2018}. In all presented experiments, the probing electrons are generated by 400\,nm laser pulses (40\,fs duration) via two-photon photoemission. The electron beam has a diameter on the sample of 80-100\,µm. The \CsVSb{} samples are cleaved under ultra-high vacuum conditions (base pressure of $2 \cdot 10^{-8}$\,mbar), transferred to the measurement chamber (base pressure of $2 \cdot 10^{-10}$\,mbar) and subsequently cooled to a temperature of 30\,K using a liquid helium cryostat. 

\section{Image analysis}
\label{sec:image_analysis}
High-quality diffraction images are taken with an integration time of 11 minutes. In order to correct for spatial inhomogeneities of the amplification process in the microchannel plate, we also record a flat-field image by measuring the diffuse reflection of electrons from a rough metal surface. In a first step, the diffraction image is divided by this flat-field image. Next, a slowly varying background is removed which stems from diffuse scattered electrons. To determine this background we cut off the main lattice spots at a perimeter of 5 standard deviations before applying a Gaussian filter (standard deviation of 100 pixels, which corresponds to 6-7 times the width of the main lattice spots). This filter preserves only the long-wavelength components in the Fourier domain, allowing us to isolate the slowly varying background intensity of the image. After subtracting the background from the original flat-field corrected image, the result was binned ($15 \times 15$ pixels) and is depicted in the right panel of Fig.\,\ref{fig:Absence_CDW_130eV}a.

\section{Dynamical LEED calculations}
\label{sec:dynamical_LEED}
\begin{figure}[!htb]
    \centering
    \includegraphics{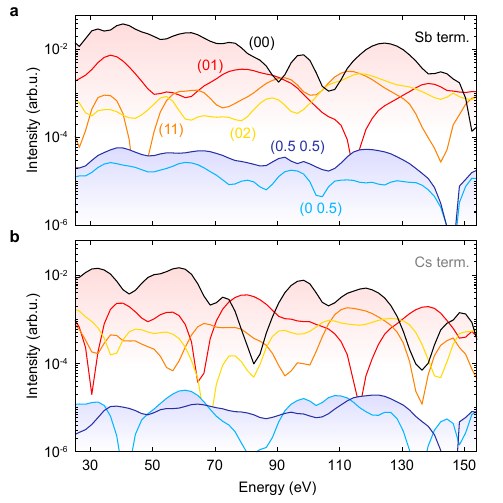}
    \caption{Dynamical LEED calculations for the hexagram distortion (instead of the tri-hexagonal distortion shown in Fig.\,\ref{fig:Dynamical_LEED}):
    Energy-dependent signal strength of the four strongest main lattice peaks as well as the two strongest CDW peaks for both Cs and Sb terminations. The distortions are taken from Ref. \cite{stahlTemperaturedrivenReorganizationElectronic2022}.}
    \label{fig:dynLEED_SoD_all}
\end{figure}

Dynamical LEED calculations were performed with the TensErLEED package \cite{blumFastLEEDIntensity2001}.
In addition to the results shown for the tri-hexagonal distortion in Fig.\,\ref{fig:Dynamical_LEED}c,d, we present in Fig.\,\ref{fig:dynLEED_SoD_all} the same curves for the hexagram distortion.

\section{Disappearance of $(\sqrt{3}\times\sqrt{3})R30^\circ$ spots}
\label{sec:sqrt3xsqrt3}

\begin{figure}[!htb]
    \centering
    \includegraphics{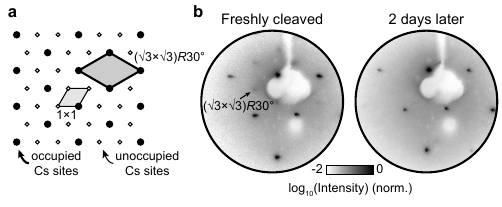}
    \caption{LEED signature of the $(\sqrt{3}\times\sqrt{3})R30^\circ$ reconstructed caesium termination.
    \textbf{a,} Schematic atomic structure of the partially occupied Cs surface.
     \textbf{b,} The $(\sqrt{3}\times\sqrt{3})R30^\circ$ spots are clearly visible after cleaving at room temperature and immediate cooling to 30\,K. They disappear after the sample was warmed up and kept at room temperature for several hours. Both LEED images of sample D are taken at a temperature of 30\,K and an electron energy of 69\,eV (left) and 70\,eV (right), respectively.}
    \label{fig:disappearance_sqrt(3)_sqrt(3)}
\end{figure}

On some surface areas, we observe additional weak $(\sqrt{3}\times\sqrt{3})R30^\circ$ spots, as can be seen faintly in the top panel of Fig.\,\ref{fig:Overview}e and more clearly in the left panel of Fig.\,\ref{fig:disappearance_sqrt(3)_sqrt(3)}b due to an optimized electron energy. These spots stem from an incomplete caesium termination, as only one third of the alkali atoms remain on the surface (illustrated in Fig.\,\ref{fig:disappearance_sqrt(3)_sqrt(3)}a) when the sample is cleaved at room temperature, as revealed by STM studies \cite{chenRotonPairDensity2021,yuEvolutionElectronicStructure2022}. Furthermore, we report on the disappearance of this caesium signature on the surface if the sample is kept for several hours at room temperature, likely due to increased Cs desorption (right panel of Fig.\,\ref{fig:disappearance_sqrt(3)_sqrt(3)}b).

\clearpage

\bibliography{CsV3Sb5_ULEED}

\end{document}